# Comparing land- and skyscapes in the three main manorial-conquered lands of the Canary Islands

María Florencia Muratore[1,2], Alejandro Gangui[1,3], Juan Antonio Belmonte[4], Carmelo Cabrera[5]

[1]CONICET - Universidad de Buenos Aires, Instituto de Astronomía y Física del Espacio (IAFE), Argentina
[2]Universidad Nacional de Luján, Departamento de Ciencias Básicas, Buenos Aires, Argentina
[3]Universidad de Buenos Aires, Facultad de Ciencias Exactas y Naturales, Argentina
[4]Instituto de Astrofísica de Canarias, La Laguna, Tenerife, España
[5]Agrupación Astronómica de Fuerteventura, Fuerteventura, España

**Abstract**

This work is a study of the relationship between astronomy and landscape focused on the orientation of Christian churches of the three main Manorial (*Señorío*) Islands of the Canary archipelago (Spain): Lanzarote, La Gomera and Fuerteventura. As a background, we have the information provided by the texts of early Christian writers, which imposed that churches should be oriented towards the east.
We carried out a comparative study between these islands to verify if the orientation patterns of the temples keep any relationship with each other, or with those of the churches of continental Europe. We are interested in exploring to what extent the indications of the early texts on Christian architecture were respected and to what degree the temples are eventually oriented following different alignments, for example according to pre-existing aboriginal traditions. We are also interested in knowing if there exist churches that are oriented towards points of the horizon on which the Sun rises on the day of the patronal feasts, since that custom was suggested in several previous studies. The analysis of the few cases in which this calendrical coincidence was verified in Lanzarote and La Gomera, is now increased with half a hundred churches that were measured in Fuerteventura.
The fieldwork that supports our comparative study is based on the measurement of the precise location coordinates, axis' azimuth and angular height of the horizon for most of the churches of the three islands, which amounts to about 120 sets of measurements. For the study of the sample, we have employed various analyses, both statistical, as well as calendric and orographic.
Our results show that on all the islands, the pattern of double orientations is repeated, which contemplates the canonical tradition of orienting the altars of churches within the solar range (pointing either eastward or westward). Very few cases also occur where it is possible to identify constructions whose orientation follows solstitial patterns, perhaps as imitation of aboriginal worship. But this double pattern also includes a high proportion of churches with orientations far from this range. An example is Lanzarote and Fuerteventura, both islands subjected to the same flow of the prevailing trade winds in the region, but each with its own characteristics. Another example is given by the particular orography of deep ravines of La Gomera, which determines the orientation of the temples located in those geographical accidents.
In this paper we show how the combination of elements of the land- and skyscape can, with a high degree of probability, offer a satisfactory explanation to the particular orientation of these insular centres of worship, which were built during the first decades after the European conquest.

**Resumen**

Este trabajo es un estudio de la relación entre astronomía y paisaje centrado en la orientación de las iglesias cristianas de las tres Islas de Señorío principales del archipiélago canario (España): Lanzarote, La Gomera y Fuerteventura. Como antecedente tenemos la información que nos brindan los textos de los primeros escritores cristianos, que prescribían que las iglesias debían estar orientadas hacia el oriente.
Realizamos un estudio comparativo entre estas tres islas para comprobar si los patrones de orientación de los templos guardan alguna relación entre sí, o con los de las iglesias de Europa continental. Nos interesa explorar en qué medida se respetaron las indicaciones de los primeros textos sobre arquitectura cristiana y en qué medida los templos se orientan eventualmente siguiendo diferentes alineaciones, por ejemplo, según tradiciones aborígenes preexistentes. También nos interesa saber si existen iglesias que estén orientadas hacia puntos del horizonte por los que sale el Sol el día de las fiestas patronales, ya que esa costumbre fue sugerida en varios estudios previos. El análisis de los pocos casos en los que se comprobó esta coincidencia calendárica en Lanzarote y La Gomera se amplía ahora con el medio centenar de iglesias que se midieron en Fuerteventura.

El trabajo de campo que sustenta nuestro estudio comparativo se basa en la medición precisa de las coordenadas de ubicación, acimut del eje y altura angular del horizonte para la mayoría de las iglesias de las tres islas, lo que suma alrededor de 120 conjuntos de mediciones. Para el estudio de la muestra se han empleado diversos análisis, tanto estadísticos como calendáricos y orográficos.

Nuestros resultados muestran que en todas las islas se repite el patrón de dobles orientaciones, que contempla la tradición canónica de orientar los altares de las iglesias dentro del rango solar (hacia el oriente o hacia el occidente). También se dan muy pocos casos en los que es posible identificar construcciones cuya orientación sigue patrones solsticiales, tal vez como imitación del culto aborigen. Pero este doble patrón también incluye una alta proporción de iglesias con orientaciones alejadas del rango solar. Un ejemplo son Lanzarote y Fuerteventura, ambas islas sometidas al mismo flujo de los vientos alisios predominantes en la región, pero cada una con sus propias características. Otro ejemplo lo da la particular orografía de los profundos barrancos de La Gomera, que determina la orientación de los templos situados en esos accidentes geográficos.

En este trabajo mostramos cómo la combinación de elementos de los paisajes terrestre y celeste puede, con un alto grado de probabilidad, ofrecer una explicación satisfactoria a la particular orientación de estos centros de culto insulares, que se construyeron durante las primeras décadas posteriores a la conquista europea.

*Palabras clave*: Islas Canarias, iglesias, orientaciones, arqueoastronomía, orografía.

**Brief history of the Canary Islands**

The Canary archipelago is made up of eight islands and numerous islets of volcanic origin, located in the Atlantic Ocean off the Maghreb coast at the point where the Sahara Desert reaches the ocean. Thanks to their orography and to the trade winds, the islands are not an insular extension of the desert. The high altitude in some areas (e.g., in Tenerife and La Gomera) and the humidity-laden trade winds that blow from the north-east make the windward areas of the islands very humid and rainy. The opposite is the case in the easternmost islands, as well as in the southern regions of most of the archipelago, where the climate is much more arid. This is the case of two of the territories we will be analyzing: Lanzarote and a large part of Fuerteventura.

The islands have been populated since ancient times. The first Berber-speaking settlers arrived from neighboring Africa between the 1st century BC and the 1st century AD. After centuries of probable isolation, the archipelago was rediscovered and visited by several expeditions from the main Italian maritime republics, e.g., Genoa, which, due to the strangulation of trade routes with the Near East by the Muslims, had embarked on oceanic navigation in search of reaching the East. Among these was that of the Genoese sailor Lanzaroto Malocello, who would arrive and occupy the island that would go into cartography with his name, Lanzarote, between 1312 and 1336 (Porro Gutiérrez 2000). Over the years, several European nations extended their Atlantic explorations, and many visited the Canary archipelago, which was finally conquered and colonized by the Crown of Castile during the 15th century. The conquest of the easternmost Canary Islands began in 1402, was led by Jean de Béthencourt and Gadifer de La Salle and authorized by King Henry III of Castile. The chronicle of this process and the start of Christianization received the name of *Le Canarien* (Aznar *et al.* 2003). After arriving and settling in Lanzarote, the expedition made incursions into the neighboring island of Fuerteventura. In 1404, Béthencourt and de La Salle founded Betancuria. Years later, a second phase of conquest took place, called the Castilian noble conquest, carried out by Castilian nobles, such as Hernán Peraza "the Elder", who appropriated the first islands by means of purchases, cessions, and marriages, and incorporated La Gomera around 1450.

These three islands are the main Manorial Islands of the Canary archipelago and are the territories in which we are carrying out our comparative study on the relationship between astronomy and landscape in terms of the orientation of the colonial churches.

**The orientation of Christian churches**

From ancient texts, including those of authors such as Origen, Clement of Alexandria and Tertullian, we know that the spatial orientation of historic Christian churches is one of the outstanding features of their architecture, with a notable tendency to orient their altars within the solar range (e.g., Vogel 1962,

McCluskey 2015). Namely, the main axis of the church, from the narthex to the altar, should be aligned with the points on the horizon from where the Sun rises on different days of the year. Among these days, there is a marked preference for those corresponding to the equinoxes (González-García and Belmonte 2015).

Although researchers have focused on analyzing specific churches in the British Isles and continental Europe, concentrating on their orientations and illumination events, studies on the orientation of temples in periods after the Middle Ages and in regions far from the European center have only gradually been developed. This late interest may be due to the widespread belief that the orientation of churches lost its importance after the 16th century (Arneitz *et al.* 2014). It is in this context that we place the present study. As we shall see, a large majority of the churches on the three islands we will consider in this work began to be erected decades after the conquest and colonization of these territories by the Norman and Castilian noblemen, who had the approval and support of the Crown of Castile during the 15th century (Cioranescu 1987).

**Historic churches of the main Manorial Islands**

Religious architecture on these islands began with the building of modest chapels with a single room. In some of them, over time, small shrines or altars were added in their headers, together with vestries on their sides and other elements of practical use, such as low walls bordering the atrium (as in San Sebastián, in Lanzarote) and calvaries or small crosses (as in Nuestra Señora del Buen Viaje, in Fuerteventura; Figure 1). Nowadays, small chapels are in general located far from the cities, while some of the churches located in the most populated areas eventually achieved some monumental dimensions.

We will now describe the particularities of the different groups of churches -and the orientation patterns that were found for them- in each of the three islands that were surveyed by our team. Then, in the penultimate section, we will present a brief comparative study to emphasize how the difference in the landscape among the islands helped in explaining some of the particular orientations that were found.

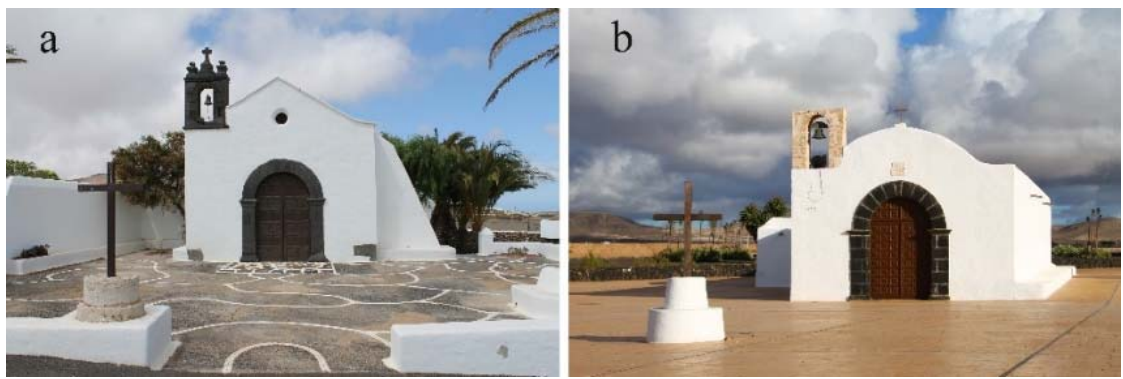

Figure 1. Simple and modest chapels in Lanzarote and Fuerteventura: (a) San Sebastián, in El Mojón, and (b) Nuestra Señora del Buen Viaje, in El Cotillo.

**Lanzarote: the imprint of the trade winds**

The churches of Lanzarote generally welcome the parishioners with a sober semicircular arched doorway marked out in local stone and, sometimes, there is also a second doorway on one of its sides. They have a modest belfry with one or more openings and a gabled or hipped roof, plastered or tiled. They are often surrounded by an exterior wall plastered in white and a barbican, simple or more prominent, located at one of its ends, presumably to cut off the wind. Inside, the chapels may have their nave and altar at different heights and be covered with simple Mudejar-style wooden ceilings, in troughs or with four or at most six gables, often with strapwork.

The orientation of the churches of Lanzarote was studied by Gangui *et al.* (2016) and, in Figure 2a, we show the diagram of azimuths of their principal axis, from the front door towards the altar. The diagonal lines on this graph indicate the extreme values of the corresponding azimuths for the Sun (continuous lines for the solstices) and for the moon (dotted lines, indicating the position of the major lunistices).

In Lanzarote, 32 churches were measured, with 15 in the solar range (Figure 2a). We can distinguish two distinct orientations: (i) to the north, with "entrance" on the leeward side, avoiding perhaps the dominant winds of the place, and (ii) eastward, with the apse of the church pointing toward the eastern sector. It seems to be a case where practical issues (the orientation against the trade winds from the N-NE; Figure 2b) appear side by side with cultic and canonical traditions (i.e., the orientations within the solar range). This pattern of orientation may reflect the desire to avoid the strong winds (and the sand) prevailing on the island, which come precisely from that direction.

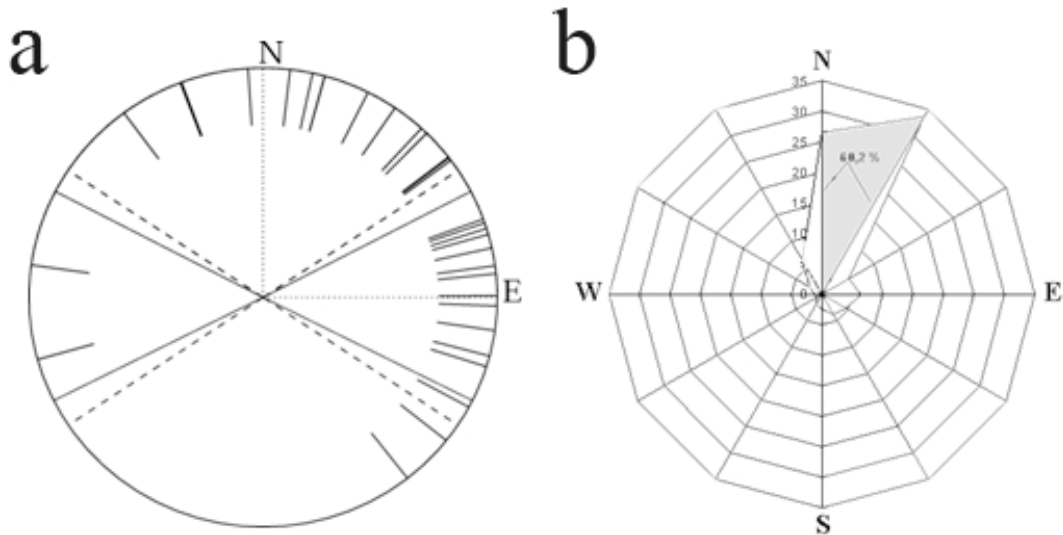

Figure 2. (a) Orientation diagram for the churches of Lanzarote. A significant number follows the canonical orientation pattern in the solar range. However, a non-negligible number are oriented to the N-NE. (b) Diagram of winds for Arrecife Airport, illustrative of the prevailing winds on the island. Note the concentration in the N-NE range, similar to the exceptional orientations of several churches on the island (Gangui *et al.* 2016).

**La Gomera and its abrupt geography**

The island of La Gomera has an abundant religious architectural heritage dating from the 16th century onwards. As in Lanzarote, the first Christian churches were small, simple chapels that were built in different areas of the island as colonization progressed. Some were located within the nascent urban centers. Others were instead founded in more peripheral locations, such as San Isidro in Roque Calvario (Alajeró), on top of the Tagaragunche mountain (Figure 3a).
The orientation of Christian churches of La Gomera was studied by Di Paolo *et al.* (2020). The corresponding diagram of azimuths is shown in Figure 3b. Of the 39 orientations measured, 12 are located in the solar range, either to the east (7) or to the west (5), representing less than a third of the total. The 7 churches pointing to the east have the particularity of being among the oldest of the island. Given that this anomalous orientation pattern was neither related to the rising Sun nor triggered by the trade winds, as was the case in Lanzarote, the authors searched for other possible -practical- influences that could explain the results. Due to the particular characteristics of the island, the landscape (topography) was naturally the best guess. It turns out that many groups of churches "copy" the direction of (i.e., their main axes are locally parallel to) the deep ravines in which they are located (e.g., Valle de Hermigua and Valle Gran Rey; Figure 3c). In these two cases the ravines follow a SW-NE line, coinciding with the accumulation of orientations in the NE region of the diagram in Figure 3b. So, this suggests that in La Gomera the accumulation of orientations in the northeast sector of the azimuth diagram is most probably due to the orographic characteristics of the island.

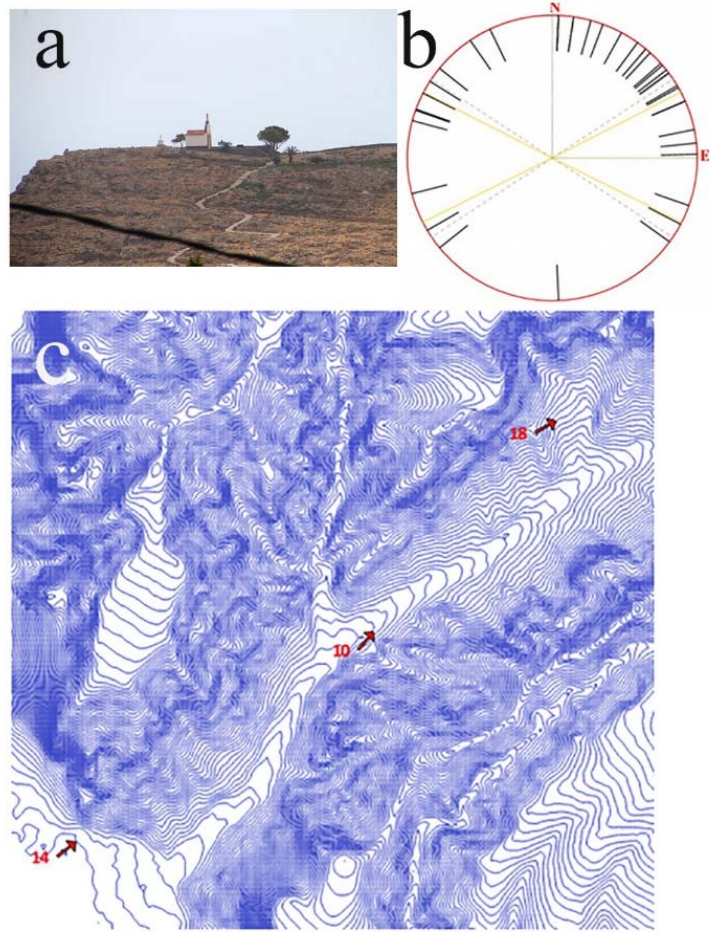

Figure 3. (a) The small chapel of San Isidro, in Roque Calvario, at the top of a mountain. (b) Orientation diagram for the churches of La Gomera. Although several buildings -about twelve- follow the canonical orientation within the solar range, a large number is aligned to the northeast. (c) Topographic map for the Valle Gran Rey and the orientation of some of the churches (small arrows) towards northeast due to the orographic characteristics of the island (adapted from Di Paolo *et al.* 2020).

**Fuerteventura: hints of Easter and stellar orientation?**

The first religious constructions on Fuerteventura were small, simple chapels with a single enclosure. One example is Nuestra Señora de Guadalupe, built in 1642 in the village of Agua de Bueyes, in the central part of the island (Figure 4a). This church, like so many others, has a rectangular ground plan, a flat front façade and is now covered by a three-slope tile roof. Many of these chapels have more than one entrance door, such as Guadalupe, whose side entrance is open towards the southern sector.
The pattern of the orientation of the churches of Fuerteventura was studied in (Muratore *et al.* 2023). In Figure 4c we present the declination histogram (or curvigram) for the sample of 48 churches, with the normalized relative frequency, which enables a detailed view of the probability density of certain orientation patterns that might be a matter of interest. The plot shows a preference for orientations within the solar range. Moreover, two peaks in declination stand out above the 3σ level. The principal maximum appears at a declination of c. 5°, while a second one, almost as significant as the former, is located at c. –14°. These peaks have a priori no obvious reason to stand out from the rest of the declination measurements that fit within the solar arc. The usual theory of orientations toward the sunrise on the church patron's feast day was discarded by individual evaluation as no patron saint's day fits with the corresponding astronomical orientation.

Muratore *et al.* (2023) offer three possible explanations to try and justify the peak of –14º in declination, although none of them is conclusive. First, the traditional Canarian celebration of "Los Finaos" (the local version of the Day of the Dead) which takes place on the night of 1 to 2 November each year. Second, a topographic orientation for churches located in a few valleys of the island pointing towards a direction close to the 105º azimuth. Lastly, an unusual but appealing orientation to a "bright star" which might be supported by some relevant ethnographic testimonies.

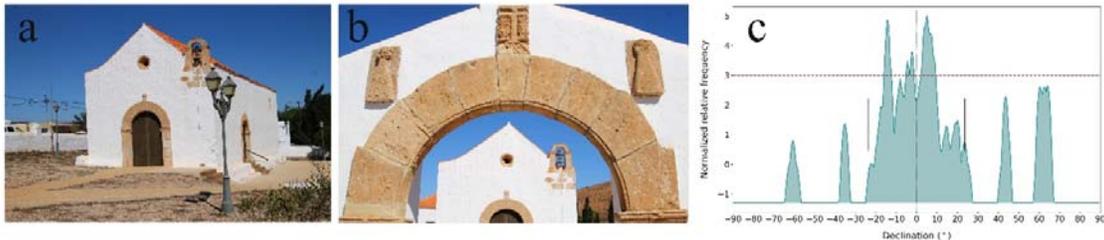

Figure 4. (a) Nuestra Señora de Guadalupe is a typical chapel of Fuerteventura. It has a secondary entrance door open towards the southern sector. (b) The chapel is surrounded by a low wall with a large arched and decorated doorway with small plaques, two of which show 8- and 6-pointed rosettes which could be interpreted as astral symbols, sometimes associated with Venus and the Sun, respectively. (c) Normalized declination curvigram of the churches. The astronomical equinox is marked with a vertical dotted line, whereas the solstices have shorter vertical solid lines. The horizontal dashed line represents the 3σ level. Peaks rising above this line can be considered as matter of interest (Muratore *et al.* 2023).

Regarding this last possible explanation, the hypothesis would be to assign this peak to a non-solar origin, perhaps prior to the Christianization of the region. It might be due to the traditional observation of a brilliant planet or a bright star. The authors propose that the *Gañanera* (the local name of Sirius), which in the 17th century had a declination of c. –16.3º (not far from –14º), is perhaps an interesting possible target. The *Gañanera* is well known in Fuerteventura as the best guide for agricultural activities, specifically, in the central area of the island where several of the churches of this group of orientations are located, in particular, the 17th century church of Nuestra Señora de Guadalupe (Figure 4a-b).
Regarding the main peak of the declination curvigram (Figure 4c) centered at c. $\delta = 5º$, Muratore *et al.* (2023) propose one possible explanation. Namely, a not negligible group of churches were not oriented to the ecclesiastical equinox on 21 March, as canonical texts apparently suggest, but to the sunrise on Easter Sunday at the approximate year of the construction of the building (as demonstrated in the Way of Saint James by Urrutia-Aparicio *et al.* 2021). Easter is indeed one of the most important feast days of Christianity. And this results in an accumulation of orientations slightly to the north of due east, as the peak in $\delta = 5º$ in the curvigram actually shows.

**Comparative study: the orientation of churches and the landscape**

The study of the astronomical orientations of these churches in their respective islands offers us the opportunity to verify whether the typical orientations found in Europe were rigidly transferred to these colonies or, eventually, whether there were influences from the pre-existing aboriginal cultures built-in the churches up to today. In general, we can affirm that most of the churches we have surveyed do not reflect pre-Hispanic traditions in their orientations. We know that the summer solstice was the most important festival for the original native people of the islands, with a concept similar to the equinox in the second term (Belmonte, 2015). The former, however, is not present in the samples of churches while the latter can easily be confused and would be indistinguishable with the ecclesiastical equinox on 21 March. Therefore, the results analyzed here are inconclusive in this particular aspect.
As we reviewed above, the orientation pattern of many churches in Lanzarote was influenced by the strong trade winds ubiquitous in the region. In fact, the areas where more churches facing north-northeast have been built (with their entrance oriented towards the southern quadrant) is on the verge of El Jable (north and center of the island), where it becomes imperative to avoid the sand driven by the wind.
In La Gomera it is possible to conjecture that many of the first churches on the island were oriented in the solar range (although this was not a rule), but that over the centuries, in the more modern churches, what prevailed was the adaptation to the orography of their sites. This can be clearly observed in the ravines of Hermigua and Valle Gran Rey (Figure 3). On the other hand, here the winds do not seem to have been as much of a conditioning factor for the orientation of the churches as they were on Lanzarote. These churches have no additional walls, for example barbicans, to protect the buildings from the wind, as we did see on the island of Lanzarote.

Finally, the results for Fuerteventura, although consistent with the pattern of orientations of groups of churches from earlier periods typical of the places of origin of the colonizers (Vogel 1962), differ notably from the outcomes obtained in the other islands of the archipelago. The analysis of Muratore *et al.* (2023) shows that there is a group of historic churches, mainly located in the central part of Fuerteventura, whose orientations are responsible for an unusual pattern, and which could be related to either one of the following explanations: the traditional Canarian celebration of "Los Finaos" at the beginning of November, a topographic orientation along the central valleys of the island and, lastly, an accumulation of orientations pointing to the declination of Sirius, known as "la Gañanera" by the old farmers of the island. All these hypotheses were considered by the authors and found equally speculative to a certain degree. Their study also found evidence suggesting that many of the churches were oriented towards sunrises during the wandering feast of Easter, an important festivity of Christianity. This was highlighted by a clear accumulation of orientations slightly to the north of due east. Bearing in mind that this tradition of orientations is present in medieval churches in the Castilian region of the Iberian Peninsula and given that Fuerteventura was colonized by the Crown of Castile, this result was not surprising.

**Conclusions and future directions**

In this work we showed how the landscape and the skyscape considered together were able to explain many of the resulting patterns of orientations of the colonial churches in three of the islands of the Canary archipelago. As we discussed, in all territories surveyed the canonical orientation following the rising Sun was detected, but this by its own was not enough to justify the bulk of the measurements. Each island showed its particular characteristics, related to climatic aspects (the winds) or to the orography, and these were key to make a comparative study and to appropriately explain similarities and differences among them. In the future, these studies ought to be continued with the churches on the few islands that have yet to be explored, e.g., Gran Canaria, whose collection of historic churches totals several dozen. This will allow us to complete our main archaeoastronomical project in the Canaries.

**Cited Bibliography**